\documentclass{article}
\usepackage{spconf,amsmath,graphicx, booktabs, enumitem}
\usepackage{pifont} 

\usepackage{amsmath}
\DeclareMathOperator*{\argmin}{argmin}
\title{LEARNING FROM TAXONOMY: MULTI-LABEL FEW-SHOT CLASSIFICATION FOR EVERYDAY SOUND RECOGNITION}
\name{
 Jinhua Liang$^{\star}$ \qquad 
Huy Phan$^{\star \dagger}$ \qquad 
Emmanouil Benetos$^{\star \dagger}$
}
\address{
$^{\star}$ Centre for Digital Music, Queen Mary University of London, UK \\ 
$^{\dagger}$ The Alan Turing Institute, UK
}
\begin{document}
\topmargin=0mm
\ninept
\maketitle
\begin{abstract}
Everyday sound recognition aims to infer types of sound events in audio streams. While many works succeeded in training models with high performance in a fully-supervised manner, they are still restricted to the demand of large quantities of labelled data and the range of predefined classes. To overcome these drawbacks, this work firstly curates a new database named FSD-FS for multi-label few-shot audio classification. It then explores how to incorporate audio taxonomy in few-shot learning. Specifically, this work proposes label-dependent prototypical networks (LaD-protonet) to exploit parent-children relationships between labels. Plus, it applies taxonomy-aware label smoothing techniques to boost model performance. Experiments demonstrate that LaD-protonet outperforms original prototypical networks as well as other state-of-the-art methods. Moreover, its performance can be further boosted when combined with taxonomy-aware label smoothing.
\end{abstract}
%
%
\section{Introduction}
\label{sec:intro}
Everyday sound recognition is to classify types of sound events in a recording or an online stream. It is a core task of machine listening and involves many practical applications, such as health care~\cite{qian_overview_2022}, smart cities~\cite{dove_sounds_2022}, and bioacoustics~\cite{morfi_few-shot_2021}. While many works in the past years have succeeded in recognising sound events using large amounts of labelled data~\cite{kong_panns_2020, gong_ast_2021}, these methods are ill-suited to real-world scenarios where it takes great effort to gather sufficient amounts of annotated data for each category or there exist sound events of unknown classes in the inference stage.

Recently, some works~\cite{cheng_multi-label_2019, shi_few-shot_2020, heggan_metaaudio_2022} began to research into few shot learning in audio classification. Cheng et al.~\cite{cheng_multi-label_2019} proposed One-vs.-Rest episode selection strategy and experimented their methods on the AudioSet~\cite{gemmeke_audio_2017}. Similarly, Shi et al.~\cite{shi_few-shot_2020} explored different approaches of linear regression as well as meta learning by utilising the AudioSet. Although AudioSet is a large-scale dataset containing hundreds types of sound events, it cannot be downloaded directly which makes it difficult to reproduce relevant experiments. More recently, there are some works~\cite{heggan_metaaudio_2022, wang_hybrid_2022} utilising some much smaller datasets in few-shot scenarios for fair, reproducible comparison. 
Meanwhile, Morfi et al.~\cite{morfi_few-shot_2021} collected sounds from mammals and birds. They then curated a database for few-shot bioacoustic event detection and hosted relevant challenges~\footnote{https://dcase.community/challenge2022/task-few-shot-bioacoustic-event-detection} for two years. Their work have attracted great interest from the wider AI sound community and shed light to how to improve few-shot audio classification with various deep learning techniques~\cite{li_few-shot_2022, yang_mutual_2022}. However, they restricted their research to animal sounds which form just a small number of sounds compared with generic audio taxonomy~\cite{gemmeke_audio_2017}. 

Targeting at generic few-shot audio classification, we firstly propose a new, large-scale dataset named \textit{FSD-FS} which contains 143 classes of sound events organised under a structured taxonomy. Compared with existing datasets, FSD-FS is curated for multi-label classification where different types of sound events can occur in the same sample. Based on the FSD-FS dataset, we assume that correlation among labels is helpful in a multi-label task. We thus propose a label-dependent prototypical network (LaD-protonet) to exploit the pairwise relationship between parent and children categories. Moreover, we design a taxonomy-awareness label smoothing technique to further leverage the audio ontology in the few-shot scenario. Experiments demonstrate that our proposed LaD-protonet outperforms the original prototypical network by a large margin. Compared with other state-of-the-art methods, LaD-protonet can still yield the best performance in terms of mAP, AUC, and F1-score.

The contributions of this paper are three-fold:
\begin{enumerate}[itemsep=-2pt, topsep=0pt, leftmargin=0.5cm, label=\roman*)]
    \item We curate a new, large-scale database for multi-label few-shot audio classification. It contains 143 classes under the AudioSet taxonomy. Compared with existing datasets, it is publicly available which would make it a useful database for few-shot audio classification benchmarking. The curated dataset is released on the Github\footnote{https://github.com/JinhuaLiang/LaD-ProtoNet}. 
    \item We propose label-dependent prototypical networks to exploit the relationship between parent and children classes. The proposed networks learn by converting a multi-label problem to multiple single-label classification and putting more importance on parent classes than children classes.
    \item We develop a taxonomy-aware label smoothing technique by leveraging the audio ontology of the database. Experiments show that this technique can boost model performance with a little computational cost.
\end{enumerate}

\section{RELATED WORK}
\label{sec:relate}

\subsection{Few-shot everyday sound recognition}
\label{subsec:database}
Everyday sounds cover a large range of sound events, some of which are hard to collect from nature, such as thunder and sounds from rare animals. This is how few-shot learning comes into the picture: models are trained with a few examples to recognise beyond the predefined label set. 

There exist some works applying few-shot learning to everyday sound recognition. Heggan et al.~\cite{heggan_metaaudio_2022} implemented various few-shot algorithms in some everyday sound datasets for single-label few-shot classification. Targeting the multi-label few-shot problem, Wang et al.~\cite{wang_who_2021} curated a synthesized dataset, FSD-MIX-CLIPS and FSD-MIX-SED, and compared model performance by controlling some generative factors in FSD-MIX-SED. Cheng et al.~\cite{cheng_multi-label_2019} adapted existing single-label few-shot algorithms to multi-label classification by proposing a One-vs.-Rest strategy. They then experimented their methods on the AudioSet~\cite{gemmeke_audio_2017} dataset. Shi et al.~\cite{shi_few-shot_2020} implemented meta-learning algorithms as well as linear regression on the AudioSet and found that meta learning performed better than other few-shot methods. We note that while Cheng and Shi both conducted experiments using AudioSet, their results are not comparable because AudioSet is not released to the public directly and neither of them detailed how they set up their database for few-shot learning.

This work curates a new dataset called FSD-FS to facilitate research on few-shot everyday sound recognition. FSD-FS offers several advantages over existing datasets as follows.

\textit{Class diversity}: FSD-FS consists of more than a hundred classes of sound events, facilitating research on generic audio classification.

\textit{Polyphonic sound events}: We assume that models will perform better if data points in the training set are closer to real-world ones. Different types of sound events are often simultaneous in the real world, so each example in the FSD-FS database can belong to multiple classes of sound events. 

\textit{Open resource}: The FSD-FS database is publicly released for research use. We hope this will accelerate further investigation on multi-label few-shot audio classification.

\subsection{Prototypical networks}
\label{subsec:protonet}
Prototypical networks are trained with a series of ``$N$-way $K$-shot'' problems where $N$ classes are sampled for classification and each class contains $K$ support examples. They generate prototypes by clustering embedding vectors as per their classes and make prediction by measuring the distance between query points and these prototypes. Suppose a data point $x_i\in X, i\in [0, |X|)$ where $X$ is the training set. Let $S_{k}$ be the set of support samples belonging to class $k$  ranged from $0$ to $K-1$. The prototype $a_k$ can be obtained by averaging embedded support points labelled with class $k$: 
\begin{equation}
    \label{eq:protonet_prototype}
    a_k=\frac{1}{K}\sum_{x_i\in S_{k}} f_{\phi}(x_i),
\end{equation}
where $f$ denotes an embedding mapping and $\phi$ denotes parameters of the model. The probability of class $k$ is then calculated by
\begin{equation}
    \label{eq:protonet_loss} q_{\phi}(y_i=k | x_i)=\sum_{x_i\in X} \frac{\exp(-d(f_{\phi}(x_i), a_k))}{\sum_{j} \exp(-d(f_{\phi}(x_i), a_j))},
\end{equation}
where $y_i$ is the predicted label for $x_i$, $d$ is a distance function such as $\ell_2$, dot and cosine similarity. The probability distribution over classes is used to calculate the cross-entropy loss:
\begin{equation}
    \label{eq: protonet_celoss}
    L_{CE} = -\sum_{k=0}^{K-1} p(x_i)\log q(x_i),
\end{equation}
where $p$, $q$ are distributions of ground truth and predictions. 

Although prototypical networks perform well in many applications~\cite{huang_enhancing_2022, tang_two_2021}, they are ill-suited to multi-label few-shot classification directly where ``$N$-way $K$-shot'' problems are hard to create without blocking labels irrelevant to a specific task.

\subsection{Label smoothing}
\label{subsec:label_smooth}

Label smoothing originates from the idea of knowledge distillation where soft labels are derived from one-hot ground-truth by an ensemble system~\cite{hinton_distilling_2015}. It helps models avoid over-confidence in the training process. Szegedy et al.~\cite{szegedy_rethinking_2016} simplified this technique by replacing pre-trained models with a uniform distribution.
Bertinetto et al.~\cite{bertinetto_making_2020} incorporated semantic information into the ground-truth by considering distances between different classes in a taxonomy. The distance is measured by counting the intermediate nodes between two classes. However, their method is not suitable for a hierarchical label set as it cannot embed classes from different levels of the hierarchy. This work improves this taxonomy-aware label smoothing technique by adopting a different distance measurement and uses it in the few-shot scenario.

%
%
\begin{figure}
    \centering
    \includegraphics[scale=0.5]{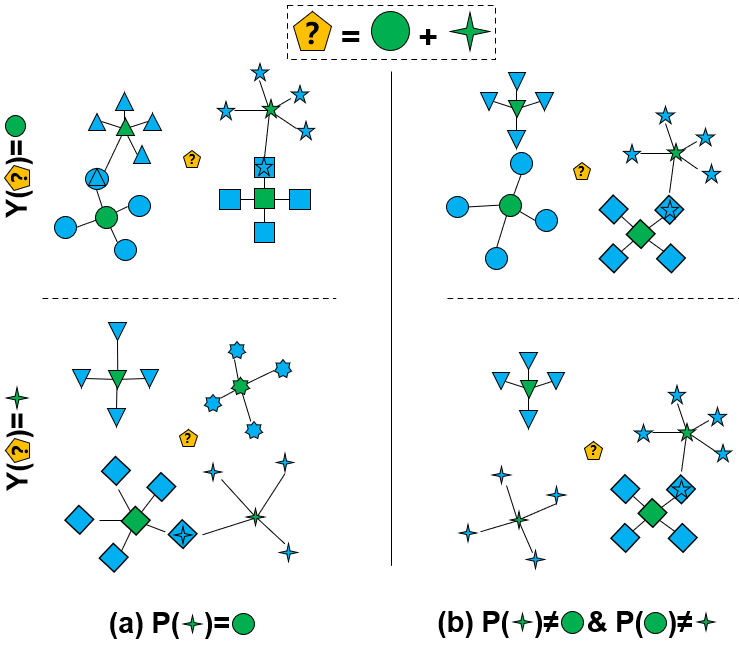}
    \vspace{-0.3cm}
    \caption{LaD-protonet in multi-label few-shot classification where there are four classes in one task and each class consist of at least four support samples. This figure shows different classes with distinct shapes. Query, support, and prototype embedding are highlighted in orange, blue, and green respectively. prototype embedding represents the classes in the latent space, calculated by averaging support embedding in each class. \textsf{Y} maps an example to its label in a specific task. \textsf{P} denotes a matching function that projects a child class to its parent class. We take an example containing two labels for examples and analyse LaD-protonet under two circumstance.  
    \textbf{Left:} these two labels are in the parent-children relationship
    \textbf{Right:} these two labels are independent with each other.}
    \label{fig:LaD_proto}
    \vspace{-0.3cm}
\end{figure}

\section{Database curation}
\label{sec:database}
We first curated our few-shot database by adapting the FSD50K dataset~\cite{fonseca_fsd50k_2022}. Compared to most of datasets for everyday sounds, FSD-50K is a database collected for multi-label classification where various types of sound events might occur in the same recording. Plus, it is an open-resource database which allows researchers to download directly. The curation of the FSD50K Few-shot (FSD-FS) dataset is detailed as follows.

1) Exclude some classes in the FSD50K dataset. FSD50K imposes the taxonomy of the AudioSet database to sort its classes. Although the AudioSet taxonomy is large enough to cover a large range of sounds in the daily life, there are some classes having multiple paths to the root which sometimes even makes human annotators get confused (e.g., ``Bicycle bell'' and ``Tuning fork'' in the AudioSet taxonomy). We thus add those multi-path classes to a black list. In addition, some levels in the tree structure do not contain enough labels for few-shot learning, so we finally maintain levels 2 and 3 out of six levels.

2) Split the label set into base, (novel) validation, and (novel) evaluation sets. Following some works in few-shot learning~\cite{alfassy_laso_2019}, we randomly split the label set by the ratio (7:2:1). We obtain a label split with 98 classes in the base set, 30 classes in the validation set, and 15 classes in the evaluation set. 

3) Adjust labels in the splits to fix smeared labels. FSD50K used smeared labels (i.e., labels propagated in the upwards direction to the root of the ontology) which could result in the increase of correlation between labels. In some cases, some combinations of labels are dominant compared to others. To tackle this problem and encourage models to learn the close relationship between labels, we slightly adjust the obtained label set in the base split to make it contain sufficient data for few-shot learning. We double check the samples to ensure that there is no overlap between the training samples and the test samples.

4) Make the labels of the base set seen to the validation and evaluation sets. Different from other multi-label few-shot audio datasets, examples in our curated dataset will be maintained for validation or evaluation if they are associated with both the base and the novel sets. We believe a good model can still recognise those audio features learned in the training process when evaluated in an unfamiliar scenario. While this work is not extended to ``few-shot learning without forgetting'' problem, the curated dataset can prompt further investigation on this problem or some more interesting, realistic directions.
%
%

\section{PROPOSED METHOD}
\label{sec:propose}

\subsection{label-dependent prototypical networks}
\label{LaD-protonet}
Prototypical networks learn an embedding space where probabilities can be obtained by computing distances over queries and the prototypical representation of each class~\cite{snell_prototypical_2017}. Prototypical networks perform well in few-shot scenarios as an episodic algorithm where learning proceeds throughout `$N$-way $K$-shot' tasks. However, the `$N$-way $K$-shot' scenario is unsuitable for multi-label classification. The reason is two-fold: (i) each label is associated with multiple classes; (ii) labels are correlated with each other. We hereby propose label-dependent prototypical networks (LaD-protonet) for multi-label few-shot classification. Its training procedure can be divided into the following four steps.

\textbf{\textit{Task formation}}:
As shown in Fig.\ref{fig:LaD_proto}, the LaD-protonet takes one example as a query point and creates support sets for each label associated with this query. Each support set consists of $N$ classes where one active class is obtained from the query point and $N-1$ classes are sampled from the rest of non-active classes. Support points are samples from examples labelled with these selected classes. In the classification, only labels included in the selected classes will remain. We refer to such a classification a \textit{task}. In this way, LaD-protonet converts a multi-label classification problem to multiple single-label classification tasks in the training stage.

\textbf{\textit{Parent-children alignment}}:
While a multi-label classification is broken down into several independent tasks in the task formation, this might jeopardise correlation between labels implicitly. Inspired by \cite{goyal_hierarchical_2021} where learning begins from abstract classes, LaD-protonet is devised to leverage the pairwise relationships between children and parent classes. When a query point is associated with both children classes and their parent classes, LaD-protonet aligns these two corresponding tasks. Suppose one example $x_i$ belongs to classes \textit{$\textbf{c}_1$} and \textit{$\textbf{c}_2$}.
Let $T_{i}^{\textit{\textbf{c}}_l}$ be the task of $x_i$ corresponding to class \textit{$\textbf{c}_l$}.
As shown in Fig.\ref{fig:LaD_proto}(b), if \textit{$\textbf{c}_2$} is the parent class of \textit{$\textbf{c}_1$}, i.e., $\textit{\textbf{c}}_2=P(\textit{\textbf{c}}_1)$, LaD-protonet aligns $T_{i}^{\textit{\textbf{c}}_1}$ with $T_{i}^{\textit{\textbf{c}}_2}$ by replacing the support points of class \textit{$\textbf{c}_1$} with ones of class \textit{$\textbf{c}_2$} and keeping the rest in the support set.

\textbf{\textit{Prototype generation}}:
Similar to original prototypical networks, LaD-protonet generates prototypes by averaging feature embeddings belonging to the same class and then predicts probabilities over classes by calculating the distance between a query point and each prototype in the embedding space. The prototype $a_k$ can be obtained by averaging embedded support points labelled with class $k$: 
\begin{equation}
    \label{eq:proposed_prototype}
    a_{k}=\frac{1}{\lvert S_{k}\rvert}\sum_{x_i\in S_k} f_{\phi}(x_i)
\end{equation}
where $f$ denotes an embedding mapping and $\phi$ denotes parameters of models. We note that eq.(\ref{eq:proposed_prototype}) is identical to eq.(\ref{eq:protonet_prototype}) when the number of examples in $S_k$ equals to $K$ for each class. The LaD-protonet learns task $T_{i}^{\textit{\textbf{c}}_l}$ by minimising the loss function:
\begin{equation}
    \label{eq:proposed_loss}
    loss_{i}^{\textbf{c}_l}=\frac{\exp(-d(f_{\phi}(x_i), a_{\textbf{c}_l}))}{\sum_{k} \exp(-d(f_{\phi}(x_i), a_k))},
\end{equation}

\textbf{\textit{Loss calculation}}: We assume that a desired classifier recognises an abstract class better than its children classes. For example, a model is supposed to predict class `Animal' more confidently than to predict its child class `Dog'. Inspired by \cite{goyal_hierarchical_2021} in which models are forced to learn more abstract classes better, the training procedure of the LaD-protonet can be depicted as
\begin{equation}
    \label{eq:max}
    \argmin_{\phi} loss = \argmin_{\phi} \sum _{x_i\in X}\sum _{\textit{\textbf{c}}_j\in y_i}\max (loss_{i}^{\textit{\textbf{c}}_j}, loss_{i}^{P(\textit{\textbf{c}}_j)}),
\end{equation}
where $X$ is the training set and $y_i$ is the ground truth associated with $x_i$ and $P$ is a function matching a child class to its parent class.

We note that the procedure in eq.(\ref{eq:max}) will be an aggregation of the task losses if an example is not annotated with parent and children labels. Therefore, it is interesting to find that the combination of original prototypical networks with One-vs.-Rest selection strategy~\cite{cheng_multi-label_2019} is a special case of the proposed LaD-protonet.

\begin{table*}[t]
\centering
\caption{Comparison of different methods in terms of mAP, AUC, F1-score with 0.95 confidence. The best results are highlighted in \textbf{bold}.} 
\label{tab:main}
\begin{tabular}{@{}cccccccc@{}}
\toprule
                    &                            & \multicolumn{3}{c}{validation set}                              & \multicolumn{3}{c}{evaluation set}                              \\
                    & $\beta$       & mAP (\%)            & AUC (\%)            & F1-score (\%)       & mAP (\%)            & AUC (\%)            & F1-score ($\%$)       \\ \midrule
Baseline~\cite{snell_prototypical_2017}            & \ding{55} & 33.02±01.04         & 83.73±0.80          & 37.32±0.75          & 34.75±1.39          & 84.81±0.97          & 39.29±1.37          \\ \midrule
                    & \ding{55} & 38.71±01.06         & 86.07±0.29          & 41.65±0.64          & 38.71±2.00          & 86.71±1.41          & 42.82±1.93          \\
one-vs.rest~\cite{cheng_multi-label_2019}         & 15                         & 38.33±0.73          & 85.88±0.29          & 41.37±0.41          & 39.82±0.64          & \textbf{87.23±0.28} & 43.47±0.53          \\
                    & 30                         & 38.68±0.97          & 85.94±0.14          & 41.68±0.24          & 39.95±1.85          & 87.20±0.60          & 43.63±1.30          \\
                    & 45                         & 39.60±1.22          & \textbf{86.11±0.42} & 42.42±0.97          & 38.89±1.38          & 86.89±0.55          & 42.77±1.11          \\ \midrule
                    & 15                         & 39.36±0.90          & 86.10±0.33          & 42.04±0.58          & \textbf{40.33±1.57} & 87.10±0.73          & 43.82±1.21          \\
LaD-protonet (ours) & 30                         & 39.71±0.56          & 85.77±0.67          & 42.16±0.98          & 40.05±0.58          & 87.04±0.58          & \textbf{43.97±0.29} \\
                    & 45                         & \textbf{39.98±0.51} & 86.01±0.30          & \textbf{42.47±0.57} & 39.68±0.75          & 86.97±0.31          & 43.40±0.74          \\ \bottomrule
\end{tabular}
\vspace{-0.1cm}
\end{table*}

\subsection{Taxonomy-aware embedded label}
\label{sec:embedded_label}

\begin{figure}
    \centering
    \includegraphics[scale=0.2]{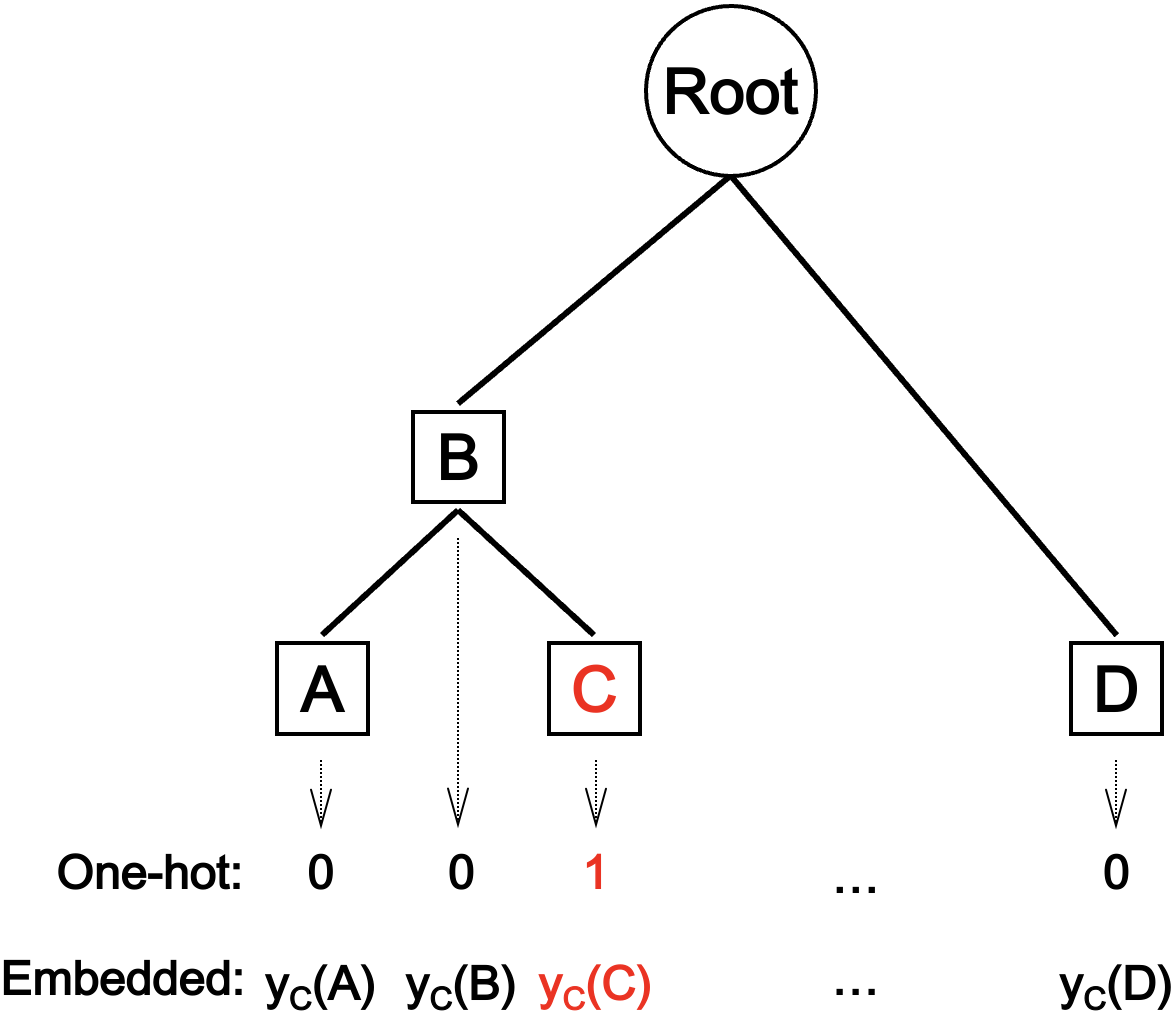}
    \vspace{-0.3cm}
    \caption{Comparison between one-hot and taxonomy-aware embedded labels. The label set~$\mathcal{L}$ consists of class $A, B, C, \dots D$. The positive class is highlighted in red.}
    \label{fig:embeded}
    \vspace{-0.3cm}
\end{figure}

In addition to pairwise correlation, we embed audio taxonomy knowledge into the soft-label ground-truth. Fig.~\ref{fig:embeded} compares the proposed taxonomy-aware embedded label with a one-hot one. In taxonomy-aware embedded labels, each negative class is assigned with a nuance probability instead of 0 while the probability of the positive class is slightly lower than 1. The probability of class $A$ in taxonomy-aware embedded labels can be calculated as follows.
\begin{equation}
    \label{eq:embed_prob}
    p_{C}(A)=\frac{\exp(-\beta d(\textit{\textbf{c}}_A, \textit{\textbf{c}}_C))}{\sum _{\textit{\textbf{c}}_C\in \mathcal{L}}\exp(-\beta d(\textit{\textbf{c}}_i, \textit{\textbf{c}}_C))}
\end{equation}
where $\mathcal{L}$ is the label set of a task and $\beta$ is a hyper-parameter controlling how class probabilities are distributed in embedded labels. The proposed taxonomy-aware embedded labels approximate one-hot labels when $\beta$ gets higher and distribute uniformly otherwise. 
%
%
\section{EXPERIMENTS}
\label{sec:exp}

\subsection{Experimental setup}
\label{subsec:exp_setup}
In all the experiments, we set the sampling rate of the audio recordings to 44.1kHz. The window length is about 20ms with 50\% overlap, and the number of Mel bank filters is 64. Log-Mel spectrograms are used as input for few-shot learning methods. Z-score normalisation is applied along each Mel bin before feeding the acoustic features into the networks. We adopt a 12-way classification scenario to match the training and evaluation process. We apply prototypical networks as our baseline system. We note that we exclude irrelevant labels whose classes are not sampled in a ``$N$-way $K$-shot'' problem.

\subsection{Network architecture}
\label{tab:subsec_architecture}

Inspired by the VGGNet architecture~\cite{simonyan_very_2015}, we design a convolutional neural network with 8 convolutional layers as the feature embedding extractor.
Each block consists of two identical convolutional filters. Except for the last block, max pooling layers with strides 2 are appended to these blocks. A global pooling operation is used in the last block. We applied this feature embedding extractor to all the few-shot learning models for a fair comparison. Details of the network architecture can be found in our released code~\footnote{https://github.com/JinhuaLiang/LaD-ProtoNet}.

\subsection{Experimental results}
\label{subsec:exp_results}

Table~\ref{tab:main} compares the performance of different methods in terms of mAP, AUC, and F1-score. Our proposed LaD-protonets with embedded labels yield better performance than the baseline system and the one-vs.rest method over all the evaluation metrics. When $\beta$ equals to 30, our LaD-protonets outperform the one-vs.rest methods in the evaluation set by 0.1\% in terms of mAP, 0.5\% in terms of F1-score, which indicates that pairwise relationships between parent and children classes help models learn useful features. In addition, the one-vs.-rest methods with $\beta$=30 yield better performance than the one without embedded labels. This is because embedded labels soften model decision by harnessing audio taxonomy knowledge in the training process. 

\begin{figure}
    \centering
    \includegraphics[scale=0.5]{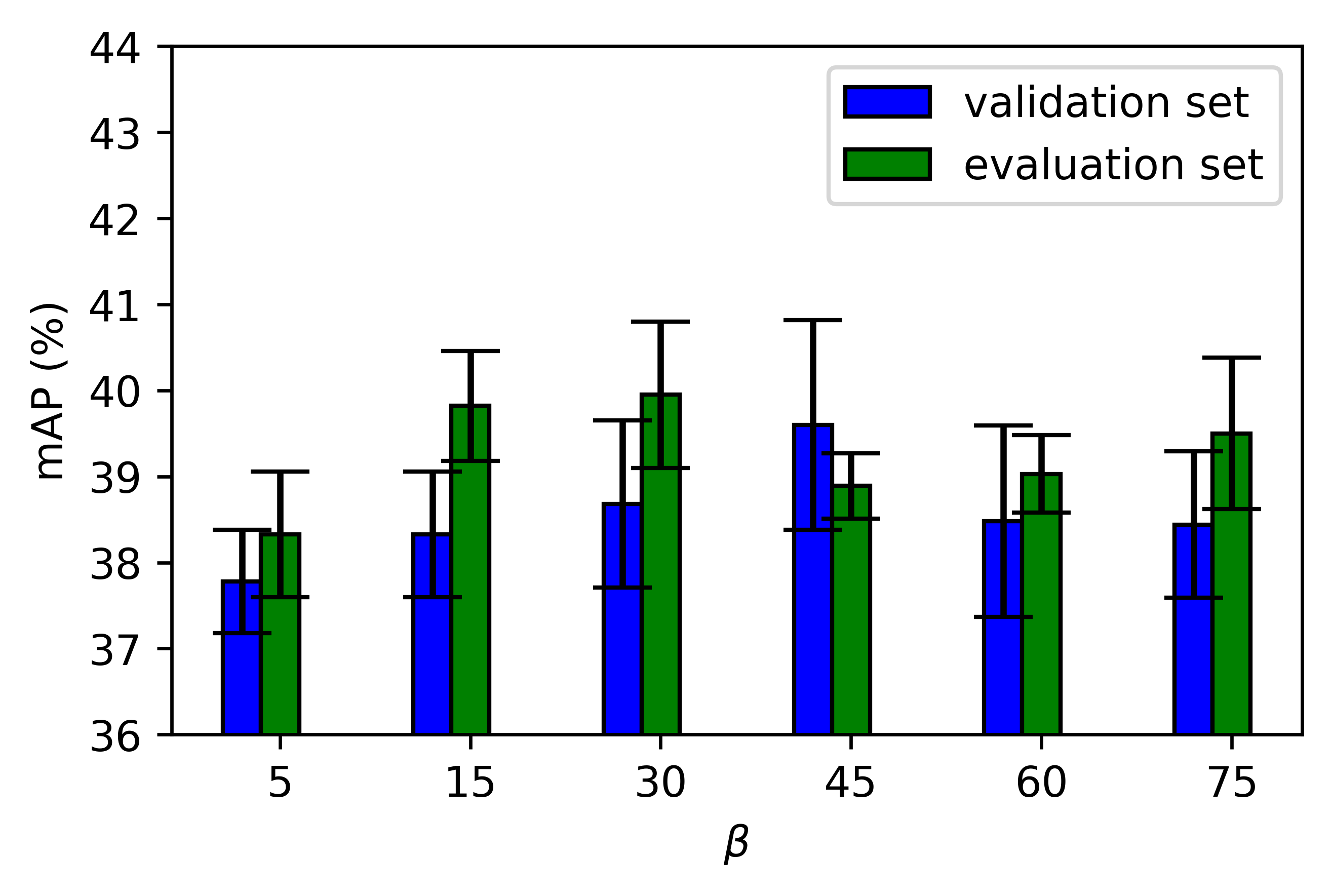}
    \vspace{-0.3cm}
    \caption{Comparison among taxonomy-aware embedded label with various $\beta$ values.}
    \label{fig:embed_result}
    \vspace{-0.3cm}
\end{figure}

Comparison among the taxonomy-aware models with embedded labels in terms of $\beta$ is shown in Fig.~\ref{fig:embed_result}. Models with $\beta$=30, 45 yield the peak performance in the evaluation and validation set respectively. We note that the model with $\beta$=45 performs worse on the evaluation set than the one with $\beta$=30 while it achieves the best performance in the validation set. This might be due to outfitting, indicating that models' generalisation decreases when labels approximate one-hot encoding format.
%
%
\vspace{-0.3cm}
\section{CONCLUSION}
\label{sec:conclusion}

This work firstly curates a new, large-scale dataset named FSD-FS for multi-label few-shot everyday sound recognition. It then proposes label-dependent prototypical networks to exploit pairwise label relationships by aligning children with parent classes and forcing models to classify parent classes better. Moreover, this work applies taxonomy-aware label smoothing techniques to embed audio taxonomy knowledge into labels. Experiments demonstrate that the proposed LaD-protonet outperforms the original prototypical networks by 1.34\% in terms of mAP on the evaluation set.

In the future, we plan to develop the curated FSD-FS by extending its label set, diversifying the data distribution, and refine the audio taxonomy by physical characteristics. It is also promising to explore approaches to leverage label correlation without predefined taxonomy in multi-label few-shot audio classification. 

%
%
\section{ACKNOWLEDGE}
\label{sec:acknowledge}
This work was supported by the Engineering and Physical Sciences Research Council [grant number EP/T518086/1]. The research utilised Queen Mary's Apocrita HPC facility, supported by QMUL Research-IT, http://doi.org/10.5281/zenodo.438045.
%
%
\bibliographystyle{IEEEbib}
\bibliography{refs}

\end{document}